\documentclass[useAMS,usenatbib,usegraphicx]{mn2e}
\usepackage{booktabs} 
\usepackage[dvips]{graphicx}
\usepackage[usenames,dvipsnames]{color}
\usepackage[pdftex]{hyperref}
\usepackage{amssymb} 
\usepackage{amsmath}
\usepackage{times}
\title{A Strip Search for New Very Wide Halo Binaries }
\date{   }
\author[Quinn ]{D.P. Quinn$^{1}$, M.C. Smith$^{1}$\\
$^{1}$ Institute of Astronomy, University of Cambridge, Madingley Road,
Cambridge CB3 0HA, UK \\
}


\begin{document}
\maketitle
\begin{abstract}
We report on a search for new wide halo binary stars in SDSS Stripe 82. 
A list of new halo wide binary candidates which satisfy common proper motion and photometric constraints is provided. The projected separations of the 
sample lie between 0.007-0.25 pc. Although the sample is not large enough to improve constraints on dark matter in the halo, we find the wide binary 
angular separation function is broadly consistent with past work.
We discuss the significance of the new sample for a number of 
astrophysical applications, including as a testbed for ideas about 
wide binary formation. For the subset of
candidates which have radial velocity information we make use of
integrals of motion to investigate one such scheme in which the origin
of Galactic wide binaries is associated with the accretion/disruption
of stellar systems in the Galaxy. Additional
spectroscopic observations of these candidate binaries will strengthen their
usefulness in many of these respects. Based on our search experience in Stripe 
82 we estimate that the upcoming Pan-STARRS survey will increase the sample size of wide halo binaries by over an order of magnitude.

\end{abstract}

\newcommand{\tot}{32644 }
\newcommand{\htot}{6419 }
\section{Introduction} 
The existence of very wide halo binaries with separations 
greater than 0.1 pc has now been firmly established 
\citep[][ hereafter CG04]{Quinn,Chan}. Such systems are intriguing 
objects for a number of reasons. 
First their formation remains a mystery. Indeed in a recent study 
\cite{Parker} show that wide binaries with semi-major axis 
$a>10^{4}$AU are too fragile to survive in low or high densities 
star cluster environments i.e. the environments in 
which most star formation is thought to take place. It is not clear if isolated 
star formation could produce wide binaries either since isolated star 
forming cores have radii of only about 0.1 pc \citep{Ward-thompson}. 
Second wide halo binaries are susceptible to disruption from 
massive compact bodies, so the distribution of wide halo binaries 
separations can be used to place constraints on the fraction of the dark 
matter halo composed of massive compact bodies \citep{yoo}. 
 
Our knowledge about wide halo binaries is still limited and 
derives mainly from the work of
\citeauthor{Chan}~(\citeyear{Chan}), who complied a list of common 
proper motion candidate binaries 
selected from the revised New Luyten Two-Tenths Catalog (rNLTT) of 
stars with proper motions greater than 180 mas/yr \citep{rNLTT,rNLTT2}. 
The sample listed 116 halo binaries pairs, with only about 10 with projected 
separation greater than $0.1$ pc. Follow up radial velocity 
measurement for 4 of these objects with separations 
greater than $0.1$ pc confirmed that 3 are likely to be 
truly associated \citep{Quinn}. 

Increasing the sample size of wide binaries is important to improve our 
understanding of these objects and quantify their implications 
for star formation and dark matter. As shown in \cite{Quinn} the existing constraints, derived from the distribution of wide halo binary separations, on 
compact massive bodies are not robust due to the small sample size. 
Larger samples are needed, moreover, to investigate the dependence of the wide 
binary separation distribution function on Galactic orbit. 
Also, with an expanded sample, the possibility of 
tracing the origin of groups of wide binaries to common 
recently disrupted Galactic satellites 
via integrals of motion \citep{Allen2} is enhanced. In addition, a number 
of other astrophysical applications, such 
as testing and constraining photo-parallax relations \citep{Sesar}, 
would benefit from increasing the sample of wide binaries. 

Here we discuss results of a search in the recently published 
SDSS Stripe 82 catalog \citep{Bramich} for new wide halo 
binaries. The outline of the paper is as follows. We first give 
an overview of the the Strip 82 survey in Section~\ref{sec:Stripe82}, 
then in Section~\ref{sec:halo} we discuss how we select halo 
stars from the survey. Section~\ref{sec:search} describes how we 
identify wide halo binaries and presents a list of new candidates. 
Finally, in Section~\ref{sec:end} we outline applications for the new sample 
and summarize our results.

\section{SDSS Stripe 82}
\label{sec:Stripe82}
The Sloan Digital Sky Survey \citep[SDSS][]{York} is a multicolour imaging and spectroscopic survey that has mapped more than a 1/4 of the sky towards the North Galactic Pole. The imaging is done with the 5 photometric bands $u$, $g$, $r$, $i$ 
and $z$ \citep{Fuk96}. A $2.5^\circ$ strip along the celestial equator (towards the South Galactic Pole) from right ascension $-49.5^\circ$ to $+49.5^\circ$, 
known as Stripe 82 has been repeatedly imaged by SDSS, from 1998 to 2005, to permit 
deeper studies and measure variability.
  
\cite{Bramich} have created a Stripe 82 light-motion catalog by 
cross-matching objects from the various imaging runs. 
The catalog contains almost 4 million light and motion curves 
of stellar and galactic objects. Each light-motion curve consists of 
around 30 epochs over a baseline of 6-7 years.
In addition a Higher-level Catalog (HLC) has been created by \cite{Bramich}\footnote{The Stripe 82 catalog including the HLC is available to download from
http://das.sdss.org/value\_added/stripe\_82\_variability/SDSS\_82\_public/}. 
This provides a range of quantities derived from the light-motion catalog, 
including mean position, proper motion and mean magnitudes in 
the five photometric bands. It is the HLC we mine in search 
of wide halo binaries.

The HLC is at present (2009) the deepest photometric and astrometric 
variability catalog of its kind and is complete down to magnitude 
21.5 in $u$, $r$, $g$ and $i$, and 20.5 in $z$. The photometric range is 14-21.5 in $r$ 
with photometric rms scatter of $20$ mmag down to $r$ $\approx 19$ which rises 
exponentially to 100 mmag at $r$=21.5. Typically proper motion errors are $\approx$4 mas/yr. \footnote{ 
We use the clipped 
photometric and astrometric measurements in the catalog which are calculated 
using an iterative algorithm that rejects the worst 4 sigma outlier in 
each iteration and terminates when there are no more outliers.}

The Stripe 82 survey provides a number of advantages over the rNLTT survey 
in the context of searching for wide halo binaries. The magnitude limits 
of Stripe 82 mean it probes much further away from the disk dominated 
solar neighbourhood than rNLTT survey (which is complete down to V=19 
and probes out to $\approx$ 250 pc from the solar position). 
Proper motion errors 
for objects in Stripe 82 are also slightly better and the 
proper motion threshold of $180$ mas/yr does not apply. Another advantage is 
that the Stripe 82 survey provides accurate photometry in 2 widely-spaced 
color bands, namely g and i, which is important 
for among other things selecting halo stars from the catalog. In contrast, errors in the V band used in rNLTT can be up to 0.25 mag. On the other hand, the rNLTT covers 44\% of the sky compared to 0.7\% for 
Stripe 82, so the density of random visual pairs is likely 
to be much higher in Stripe 82 searches.  

We note another possible approach to search for new wide binaries is 
to match the single epoch SDSS data ($14<r<20$) with older 
photometric/proper motion surveys. 
In fact a number of groups have investigated this avenue by combining 
SDSS with USNO-B \citep[][ with the latter focusing on wide disk binaries]{Chan2,Sesar}. While this approach 
covers a larger fraction of the sky, it is about 1.5 mags less 
deep than Stripe 82. 
The preliminary finding of \cite{Chan2} is 
that proper motions and photometry alone are not enough to define high probability candidates constructed from the union of SDSS and USNO-B; follow up radial velocities are needed to prune the sample. Below we show that it is possible to find a new robust sample of wide halo binaries in Stripe 82 with 
just proper motions and photometry.
\begin{figure}
\rotatebox{0}{\scalebox{.5}[.5]{\includegraphics{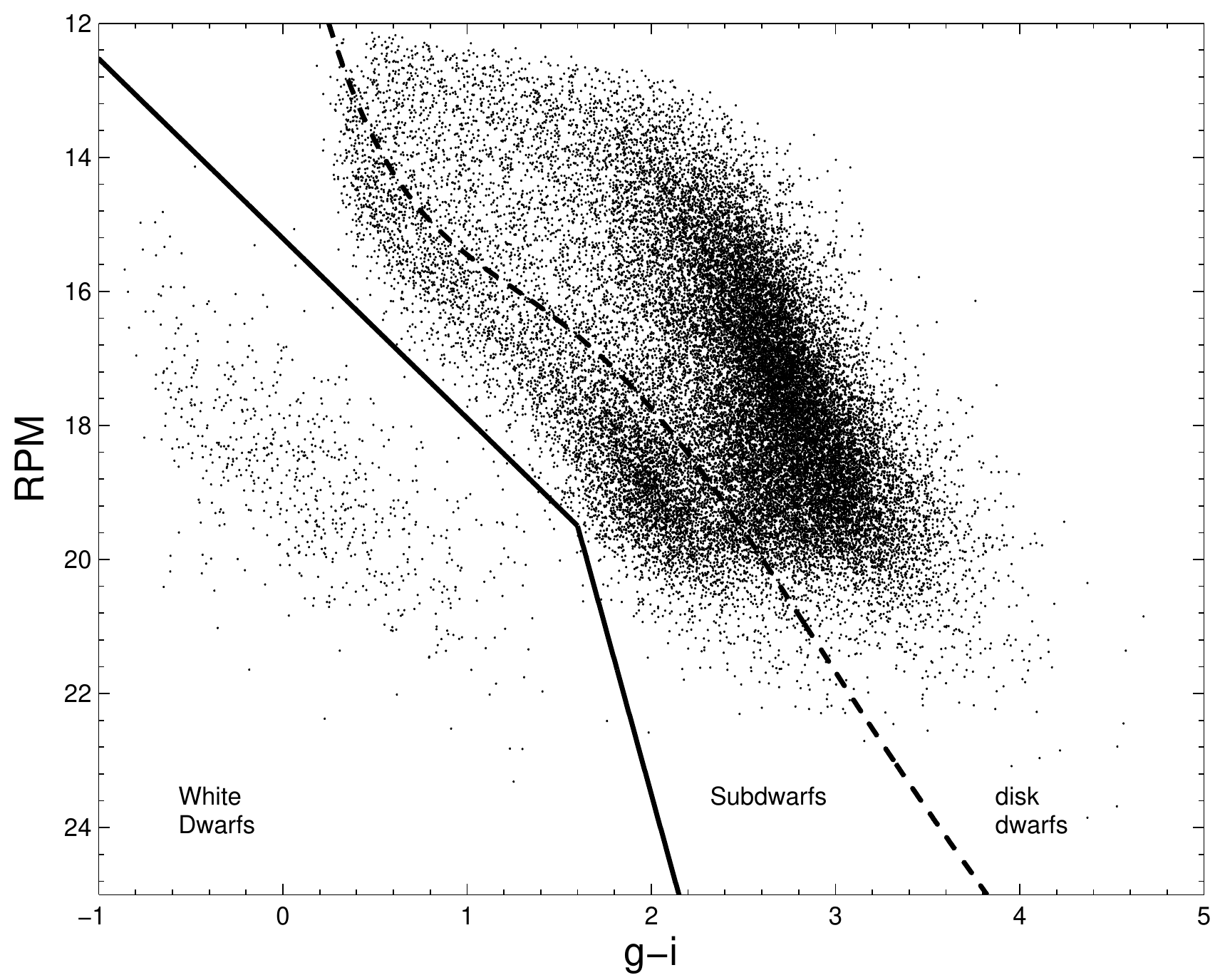}}}
\caption[]{\label{rpm}Reduced Proper Motion Diagrams  of high proper motion stars throughout the Stripe 82. The regions of the RPM diagram predominately occupied by white dwarfs, subdwarfs and disk main sequence stars are labeled and the boundaries, as discussed in the text, between the different regions are drawn.}
\end{figure}
\section{Selecting Halo Stars}
\label{sec:halo} 
The first problem to overcome in the search for halo 
binary stars from the Stripe 82 catalog is the question of how 
to select a clean sample of halo stars (subdwarfs) from the mix 
of stellar populations in the catalog. 
\subsection{ Reduced Proper Motion (RPM) Diagram}
Subdwarfs can be identified directly by their position in the HR 
diagram or features in their spectra. For the vast majority of stars in 
Stripe 82 neither spectra (but see Section \ref{sec:vrad}), nor accurate parallaxes are available.
With only proper motion and photometry on hand, the 
reduced proper motion (RPM) diagram is a common tool 
employed to pick out subdwarfs from the disk stars.  
In this diagram, the reduced proper 
motion, which we take as H$_{r}=r+5\textrm{log} \mu +5$, 
with $r$ the apparent magnitude in the $r$-band and $\mu$ the proper motion in 
arcseconds per year, is plotted against the stellar colour. 
The reduced proper motion exploits the differences 
in metalicity and kinematics between subdwarfs and disk stars. 
Since halo stars do not rotate about the Galaxy as fast as main sequence disk stars and are 
fainter at the same colour, the RPM is typically larger 
for subdwarfs compared to disk main sequence stars at a given temperature.

The colour adopted in the RPM diagram is a proxy for temperature and 
is chosen to enhance the separation of the halo subdwarfs from the disk 
main sequence stars. After some experimentation 
we found that the $g-i$ color leads to a reasonable division. This choice for colour 
and the form for H$_{r}$ have appeared before in the literature, to construct 
RPM diagrams for Stripe 82 in order to search for new white dwarf candidates \citep{Vidrih} and to study 
the kinematics of the stellar halo, (\citeauthor*{ms_tilt}~\citeyear{ms_tilt}; \citeauthor{ms_hkin}~\citeyear{ms_hkin}), 
while \cite{Sesar} constructed a similar RPM diagram from the union of 
SDSS and USNO-B across the entire SDSS footpath to define a sample 
in which to search for wide disk binaries.

Errors in proper motion measurements and also the intrinsic spread 
of velocities for a given population will mask the separation in the diagram. 
We limit the influence of proper motion errors by selecting stars 
in the catalog with proper motions larger than 40 mas/yr and with proper motion 
errors less than 5 mas/yr, the latter corresponding to a error of about 0.3 
in H$_{r}$. Errors in photometry also blur the separation in the RPM diagram. 
We require that the error of the mean photometry in the $r$, $g$, and $i$ bands to be less 
than 0.05 mags. These introduce an error in H$_{r}$ of less than 0.05 
and 0.07 in the colour. 

The reduced proper motion diagram is plotted, for the \tot stars in the HLC 
that met the
above criteria in Figure~\ref{rpm}. 
We can see that stars are found in 3 groups: white dwarfs; subdwarfs and 
disk stars. The white dwarfs to the lower left hand side of the diagram 
are clearly separated from the other groups. We use the separators put 
forward in \cite{Vidrih} to divide the white dwarfs from the rest of the stars. 
For $g-i\le 1.6$ the separator is defined as H$_{r}=2.68(g-i)+15.21$. For 
$g-i>1.6$ it is defined as H$_{r}=10(g-i)+3.5$.

It is not as clear cut where to draw the boundary between the disk main 
sequence stars and the subdwarfs, but since the 
RPM is equivalent to combining the absolute magnitude and the logarithm 
of the magnitude of the tangential velocity of the star, we can use a 
photo-parallax relation to define a boundary in the RPM diagram for
a given tangential velocity.
\citet[][ hereafter I08]{Izv}, provide a photometric 
parallax relation for stars observed in SDSS colors. The relation 
gives the absolute magnitude in the $r$-band, M$_{r}$ as a function of colour, $g-i$, and metalicity, 
[Fe/H], as follows
\begin{eqnarray}
\textrm{M}_{r} &=&-5.06+14.32(g-i)-12.97(g-i)^{2} \nonumber \\ &&+6.127(g-i)^{3}  -1.267(g-i)^{4}+0.0967(g-i)^{5}\nonumber \\ & &+ 4.5-1.1[Fe/H]-0.18[Fe/H]^{2}.\\\nonumber
\end{eqnarray}
Using this relation with the metalicity set to be -1.5 dex 
(corresponding to the median halo metalicity 
as reported in I08) and 220 km/s for the tangential velocity we 
find that, as shown in Figure~\ref{rpm}, the resultant RPM boundary 
does a good job of defining a disk/subdwarf divider consistent 
with the apparent division 
discernible by visual inspection. 
\begin{figure}
\rotatebox{0}{\scalebox{.5}[.5]{\includegraphics{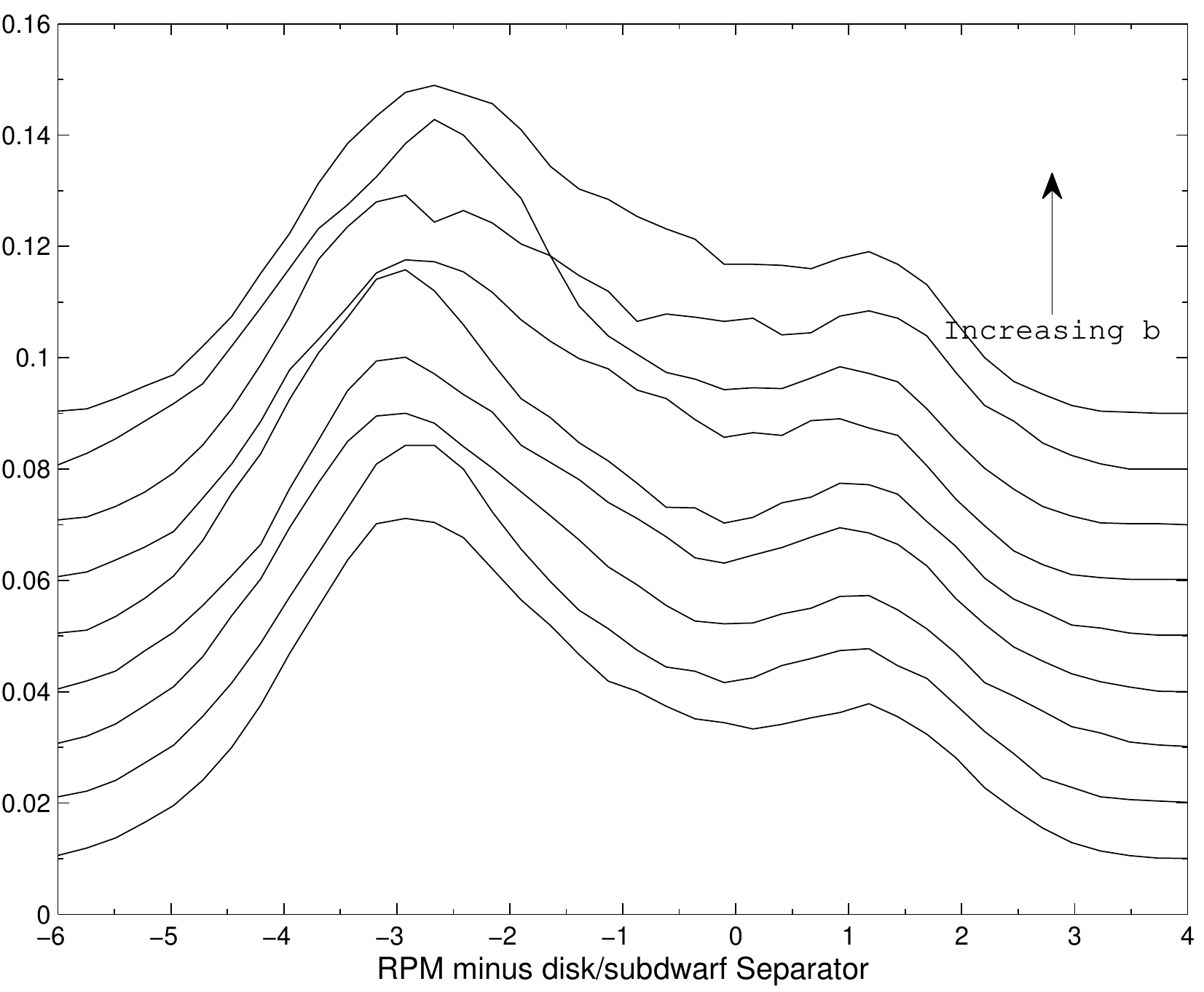}}}
\caption[]{\label{rpmdiff}
The distributions, for stars in equally spaced bins in galactic latitude, of the differences between the RPM of the star and the value of the subdwarf-disk separator evaluated at the colour of the star, are shown. A vertical offset is applied to each distribution for clarity.}
\end{figure}
CG04 made similar use of the RPM diagram to classify binaries as 
disk or subdwarfs. The definition of their disk/subdwarf separator 
contained a dependence on galactic 
latitude, $b$. We assess the need for 
such a dependence in our sample, which covers a narrower range in $b$ than the 
CG04 sample, by plotting the distribution of the differences 
between the value of H$_{r}$ and the value of our separator at the colour 
of the star for groups of stars binned in $b$. If the separator has a significant dependence 
on $b$ we can expect to see a systematic shift in the distribution of these differences as $b$ changes. 
As Figure~\ref{rpmdiff} shows there is no indication that we are detecting 
such a shift. Therefore, we define the halo subdwarf 
sample as the stars in the RPM diagram lying between the white dwarf and 
photo-parallax based divisions. This selection produces a 
sample of \htot subdwarfs.   
\begin{figure*}
\begin{minipage}[b]{0.33\linewidth}
\rotatebox{0}{\scalebox{.33}[.33]{\includegraphics{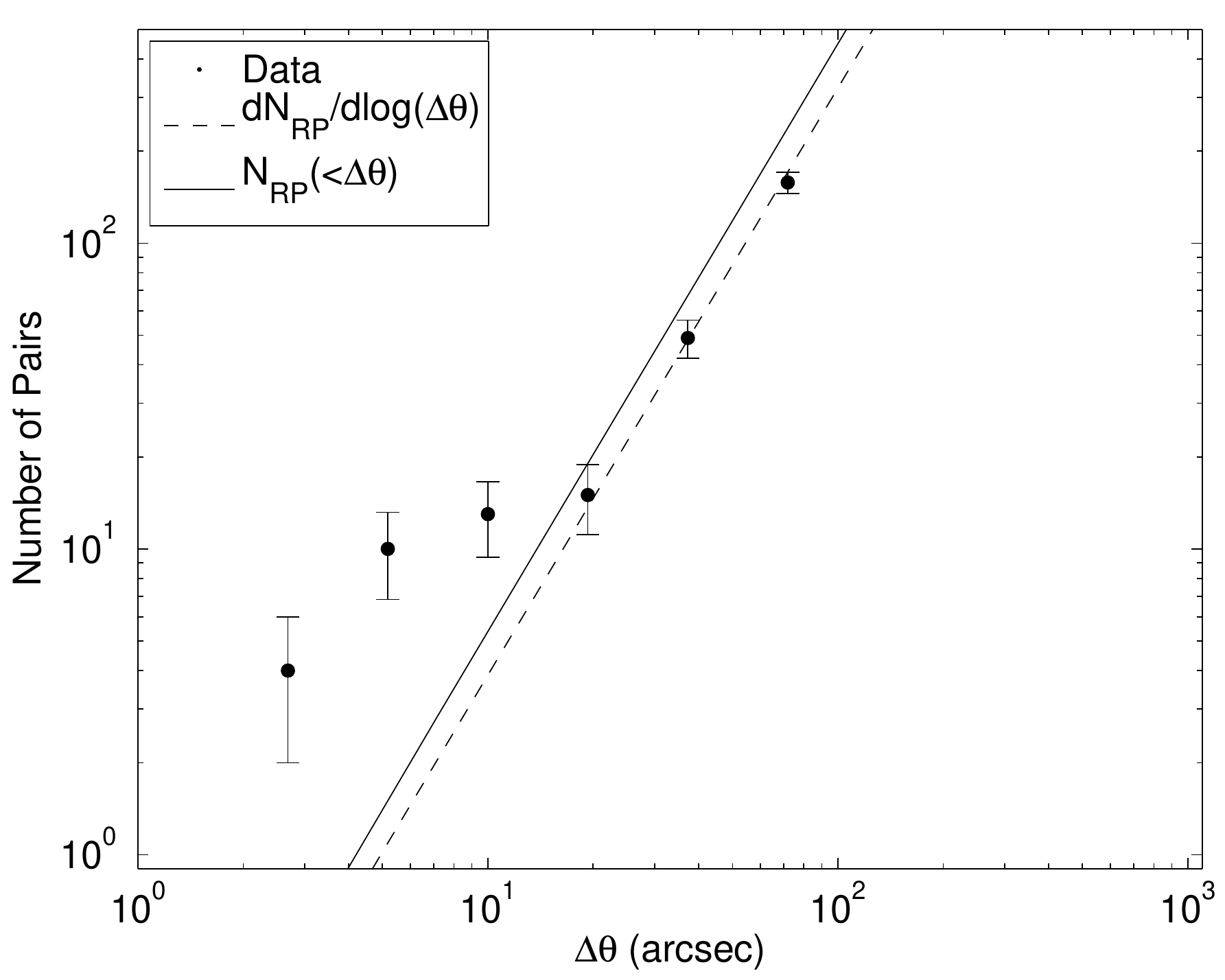}}}
\end{minipage}
\begin{minipage}[b]{0.33\linewidth}
\rotatebox{0}{\scalebox{.33}[.33]{\includegraphics{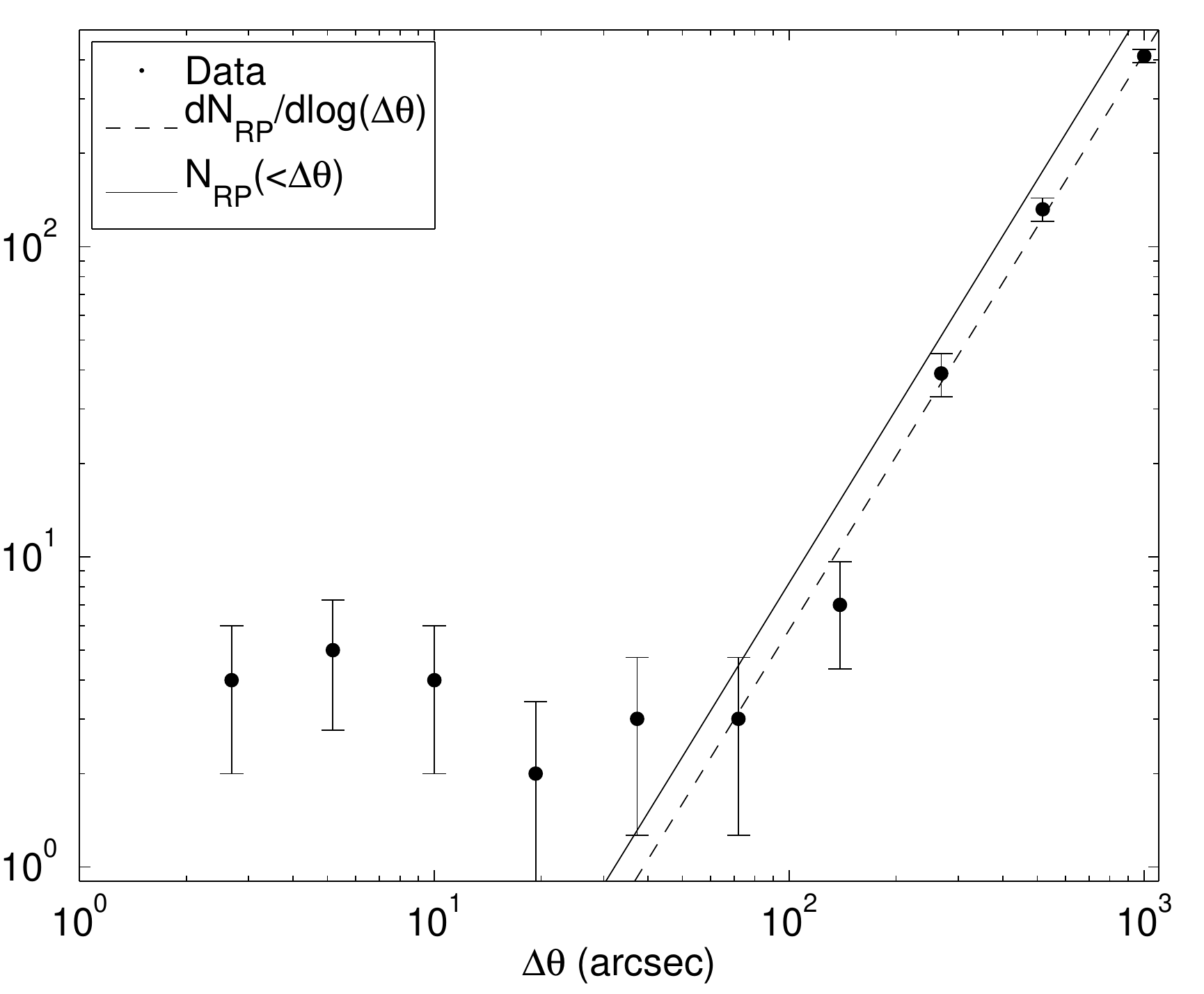}}}
\end{minipage}
\begin{minipage}[b]{0.33\linewidth}
\rotatebox{0}{\scalebox{.33}[.33]{\includegraphics{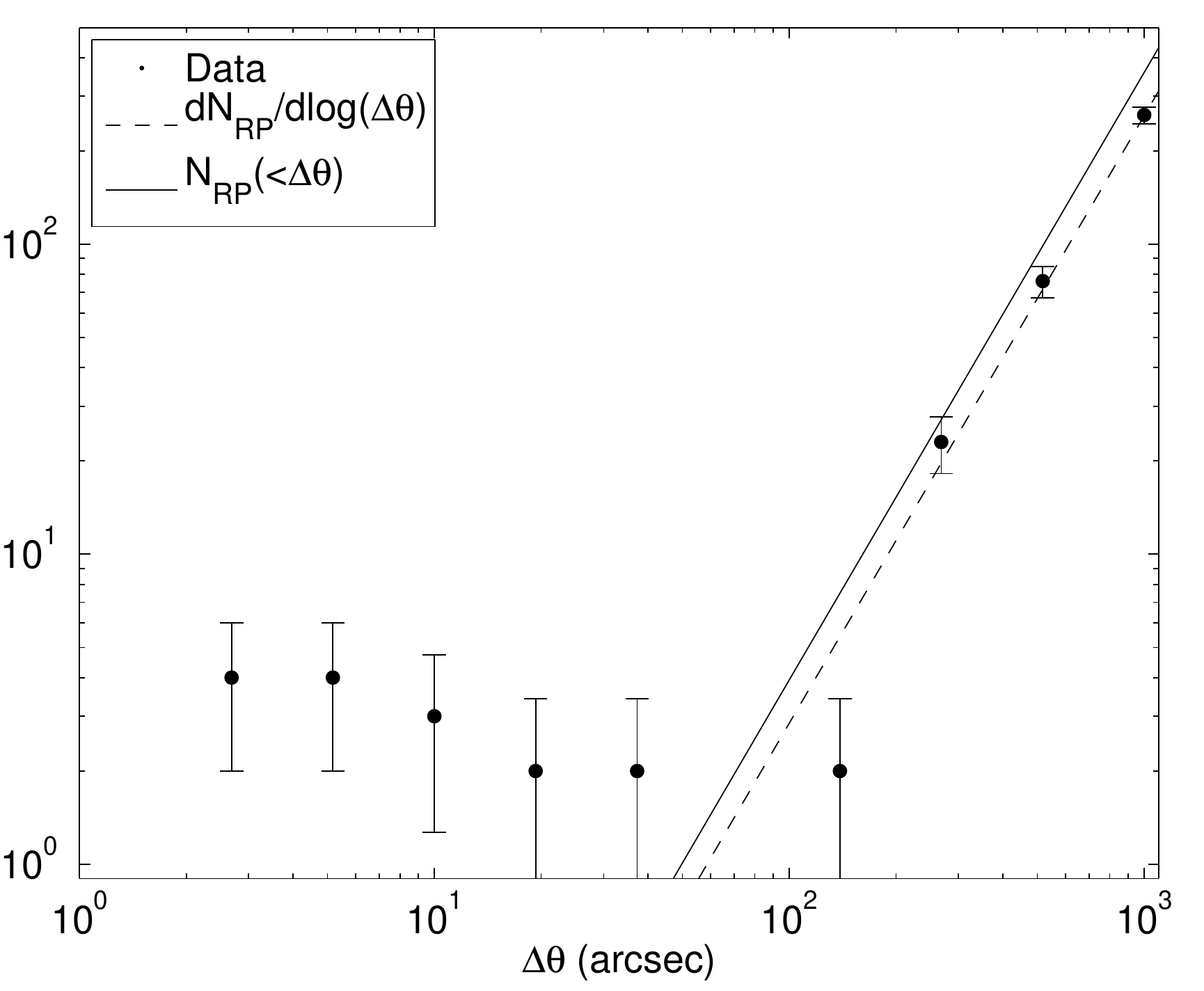}}}
\end{minipage}
\caption[]{\label{ang_halo} The distribution of angular separations 
of subdwarf pairs, selected using different criteria, 
are displayed in the 3 panels. The distributions
show the number of pairs in equal logarithmic 
bins of angular separation. 
The error-bars represent Poisson uncertainties. 
In the left panel the distribution of angular separations 
is determined by merely counting subdwarf pairs. 
In the middle panel a proper motion 
constraint is imposed on pairs. In the right panel the 
distribution is derived from pairs that meet a photometric 
parallax constraint in addition to the proper motion one. 
Also shown for each case are fits 
to the distribution of random pairs, 
$d\textrm{N}_{RP}/d\textrm{log}(\Delta\theta)$ (dashed lines)
and based on this the cumulative distribution of random pairs, 
$\textrm{N}_{RP}(<\Delta\theta)$ 
(solid lines).}
\end{figure*}
\section{Searching for Halo Binary Stars}
\label{sec:search}
With the sample of halo stars now defined, we can  proceed to search 
for binary pairs. The simplest approach is to count the number 
of subdwarf pairs as a function of angular 
separation, $\Delta \theta$, on the sky. However, this simpleminded 
approach fails to deal with problem of chance close pairs of stars. 
The contamination due to these random pairs, $\textrm{N}_{\textrm{RP}}$, 
depends on the surface density of stars in the catalog and for 
small $\Delta \theta$ we can expect roughly 
that $\textrm{N}_{\textrm{RP}}(\Delta\theta)\propto (\Delta\theta)^{2}$. 

The actual number of stellar pairs, $\textrm{f}(\Delta\theta)$, for the Stripe 82 subdwarf sample,
in equal logarithmic bins of angular separation, is shown in the lefthand panel of Figure~\ref{ang_halo}.
In order to estimate the signal from chance pairs, we 
fit a power law using a likelihood method to the distribution of 
pair separations with $100\arcsec<\Delta \theta <1000\arcsec$. We see 
this adequately fits the data beyond $\approx 10\arcsec$. 
The signal in excess of the fit inside 
$\Delta \theta <10\arcsec$ is most likely attributable 
to genuine binaries. This signal though is clearly not well 
described by a power law with an index -1.57, as observed for the 
binary angular separation function in CG04. This could be due to blending which reduces the efficiency of detecting pairs closer than $\approx 3\arcsec$ when good photometry is demanded, as noted in \cite{Sesar} and \cite{long}. 
According to the I08 relation with metalicity -1.5 dex, the subdwarf sample lie at median distance of 1.3 kpc, so the plot 
implies that the direct search for candidate wide binaries 
becomes dominated by the signal from random pairs 
at separations of around 0.05 pc and suffers from blending for 
binaries with separations less than about 0.01 pc.
\subsection{Common Proper Motion Pairs} 
Since our goal is to generate a reliable list of 
candidates rather than a complete list with many false contaminants, 
we must exploit the proper motion and photometric information to 
cut down on the signal from random pairs. 
As the typical distances to the binaries is around a kpc, 
the proper motion differences for wide binaries, arising 
from relative orbital velocity and 
projection effects due to the angular separation in the sky, 
fall below the typical proper motion error. 
Consequently, we incorporate a proper motion constraint in 
our wide binary selection procedure by requiring that 
the magnitude of the vector proper motion differences for candidate 
pairs is less than 5 mas/yr (roughly the typical relative proper motion error).
The middle panel of Figure~\ref{ang_halo} displays the outcome for f($\Delta\theta$) on imposing this constraint. 
The effect is to reduce the amplitude of the signal from random 
pairs, found using the same technique as before,
 by over an order of magnitude, without seriously compromising the 
genuine binary signal in the inner region.
\subsection{ Photometric Constraints}
We can go further and apply photometric constraints 
in the selection process. Unfortunately there are no well-defined SDSS 
colour-magnitude relations that cover the entire colour range  
for halo subdwarfs which means we cannot fully exploit the excellent Stripe 82 
photometry. The source of the problem is the fact that SDSS has a bright   
magnitude limit greater than the faint magnitude limit of the Hipparcos survey, 
which measured accurate distances to nearby bright stars permitting 
photo-parallax relations to be constructed from them.

As mentioned above, I08 presented a photo-parallax relation for SDSS 
colours. The blue end ($g-i\lesssim1$) of this relation is grounded on 
observations of stars in globular clusters with known distances. 
However, the red end has been constrained only with disk stars. 
The photo-parallax relation from IO8 should predict that 
each member of a pair have a similar value for the difference 
between the model absolute magnitude and the apparent magnitude i.e 
the quantity $\delta=M_{r1}(g-i)-m_{r1}-(M_{r2}(g-i)-m_{r2})$ should 
be centered about zero, where $M_{r1}(g-i)$ and $m_{r1}$ are the 
absolute magnitude from the I08 relation and the observed apparent 
magnitude of one of the components in the binary, and 
ditto for $M_{r2}(g-i)$ and $m_{r2}$. 
We test the relation of I08 on the the 14 pairs with 
$\Delta \theta<15\arcsec$ satisfying the common proper 
motion criterion, which we assume for the moment are all genuine binaries. 
We find the median and standard deviation in the distribution of $\delta$ for 
the binary sample turn out to be 0.04 and 0.55 mags respectively. The 
scatter is larger than the formal error in $\delta$ which is less 
than $0.1$ mags, as the median photometric errors of the 14 pairs 
are smaller than 0.01 mags in $r$, $g$ and $i$. 
The large scatter might point to the limitations of the photo-parallax relation.
 Alternatively, the wide binary sample may include a 
number with an unresolved triplet. 
We get similar scatter if we use the binary candidates
to fit a quadratic polynomial colour-magnitude relationship, $M_{r}(g-i)$.
(It turns out that attempting to fit higher order polynomials 
leads to unphysical colour magnitude relations.) 
Of course, in this approach the constant term is not constrained, so 
distances to the objects cannot be directly determined. 

We produce a final homogeneously selected sample of 
wide binary candidates by requiring that in addition to the 
proper motion constraint, $\delta$ from the I08 relation should be 
less than 0.8 mags ($\approx$ 1.5 times the scatter and corresponding 
to a distance uncertainty of around 400 pc at 1 kpc). The right hand 
panel of Figure~\ref{ang_halo} shows the improvement on applying 
this requirement. We see that the amplitude of the signal from random pairs 
falls by a factor of about 2. Wide binaries can be detected now with 
little contamination out to $\approx$ 40$\arcsec$.
In Table~\ref{tab_cand} we provide a list of the $15$ new candidate halo binaries 
with $\Delta\theta<40\arcsec$ that meet the proper motion and 
photometric constraints. We estimate that the sample contains 
about one contaminant based on the
cumulative distribution of the fit to the random pairs.  

The photo-parallax proviso removed 2 pairs with 
$\Delta \theta<15\arcsec$ that met the proper motion constraint. One of these 
which has $\Delta \theta=11.23\arcsec$
is probably spurious because the bluest star in this pair is almost 
a magnitude fainter than its apparent partner. 
The other object  has $\Delta \theta=4.54\arcsec$ with one of the components 
close to the magnitude limit in $r$. We list this object also in Table~\ref{tab_cand}, as the signal from random pairs is negligible at its angular separation (see middle panel of Figure~\ref{ang_halo}).

In an effort to boost the sample 
size further, we have considered reducing the proper motion 
threshold, but tightening the photo-parallax constraint to compensate for the  
increase in contamination from spurious pairs. However, these experiments 
failed to produce a larger sample of robust wide binary 
candidates. Lowering the threshold, as alluded to in Section 3.1, 
also runs the risk of increasing the contamination from the disk 
stars as the separation between components in the RPM diagram is 
not as well defined.
\begin{figure}
\rotatebox{0}{\scalebox{.5}[.5]{\includegraphics{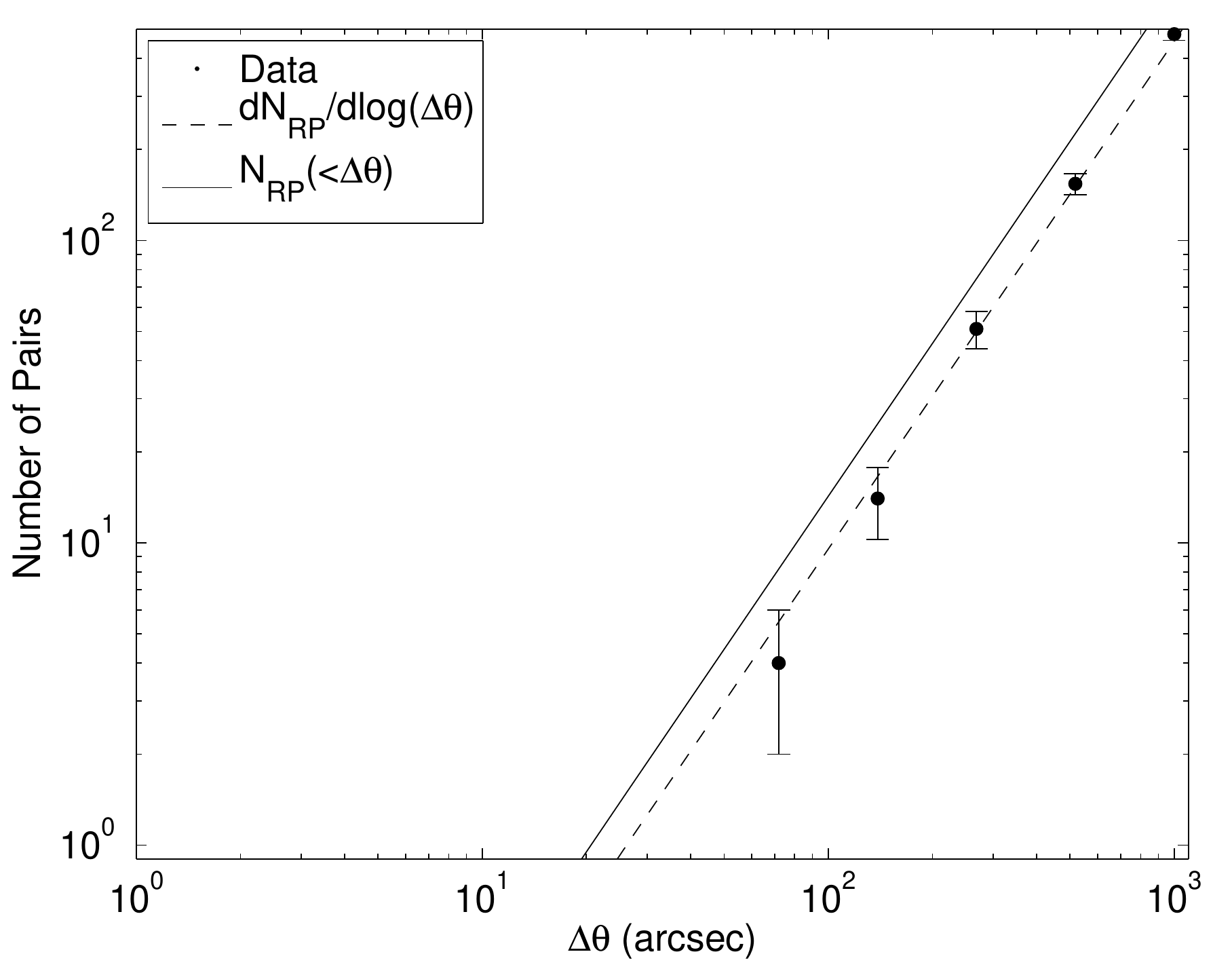}}}
\caption[]{\label{r_ang_halo}This plot shows the number 
of subdwarf pairs, similar to the left 
hand panel of Figure~\ref{r_ang_halo} except we restrict the sample 
to pairs with radial velocity measurements. Beyond an arcminute (projected separation $\approx 0.4$ pc) the distribution is well 
matched by the distribution of random pairs. The absence of pairs with separations 
less than around an arcminute is simply a reflection of the target 
spectroscopic selection effects.}
\end{figure}
\subsection{Radial Velocities in Stripe 82}
\label{sec:vrad}
We have mentioned above that SDSS is also 
a spectroscopic survey and a large sample of stellar 
spectra have been taken across the SDSS footprint \citep[][ and references therein]{Yanny_segue}, 
so it is important to investigate if we can factor 
in radial velocity or metalicity information into the binary search.
On cross matching with the SDSS catalog of stellar spectra
we find that 781 objects, about 12\% of the subdwarf sample, including 
5 stars from the homogeneous selected new wide binaries which we list in Table \ref{tab_2}, 
have radial velocities with errors less than 15 km/s. While a non-negligible
fraction, the available radial velocity information is not ideal 
for improving the detection efficiency or characterizing 
the distribution of wide binaries. This is because of a restriction 
in the SDSS spectroscopic plate which requires targets to be separated
on the sky by at least an arc-minute 
\citep[][ see also Figure~\ref{r_ang_halo}]{Yanny_segue}. So, for 
$\Delta\theta<60\arcsec$ where the signal for wide binaries is 
strongest we cannot make use of a common radial velocity 
constraint on candidate pairs analogous to the common proper motion one. 
Not surprisingly, if we nonetheless repeat the search as 
before but include a radial velocity constraint\footnote{For 
the radial velocity consistency test, a 
conservative requirement that the relative 
velocities are less than 20km/s is imposed (most of the radial velocity 
errors are actually less than 10km/s).} for any pairs in which both components have radial 
velocity measurements, we find that the fits to the distribution 
of random pairs are virtually identical to the corresponding ones in 
Figure~\ref{ang_halo}. 
We also carried out a search in the subsample of 781 objects with 
radial velocities for wide binaries. This failed to produce any candidates, 
in fact the proper motion and radial velocity constraints together 
are powerful enough to eliminate any of the potential pairs 
with separations out to 1000$\arcsec$ ($\approx 10$ pc), which 
make up the signal in Figure~\ref{r_ang_halo}.

By lowering the original proper motion threshold of 40 mas/yr, it 
is possible to get a larger sample of Stripe 82 subdwarfs with 
radial velocities. Indeed \cite{ms_hkin} have created 
a sample of 1717 subdwarfs from Stripe 82 with radial velocities 
and proper motions without imposing a proper motion threshold,
in order to study the kinematics of the stellar halo. 
We have looked also in this sample but found no high 
probability candidates. The proper motion and radial 
velocity cuts leave a pair distribution consistent with a random 
distribution of pairs. Adding a photo-parallax constraint and/or a common 
metalicity constraint just reduces the amplitude of this signal 
without revealing any excess that might be due to very wide binaries. 
 The significance of these 
null detections of wide binaries with $\Delta\theta>60\arcsec$ is not 
easy to interpret since for these samples we have no handle on number of wide 
binaries at smaller $\Delta\theta$ because of the 
spectroscopic target selection effects. 
\begin{figure}
\rotatebox{0}{\scalebox{.5}[.5]{\includegraphics{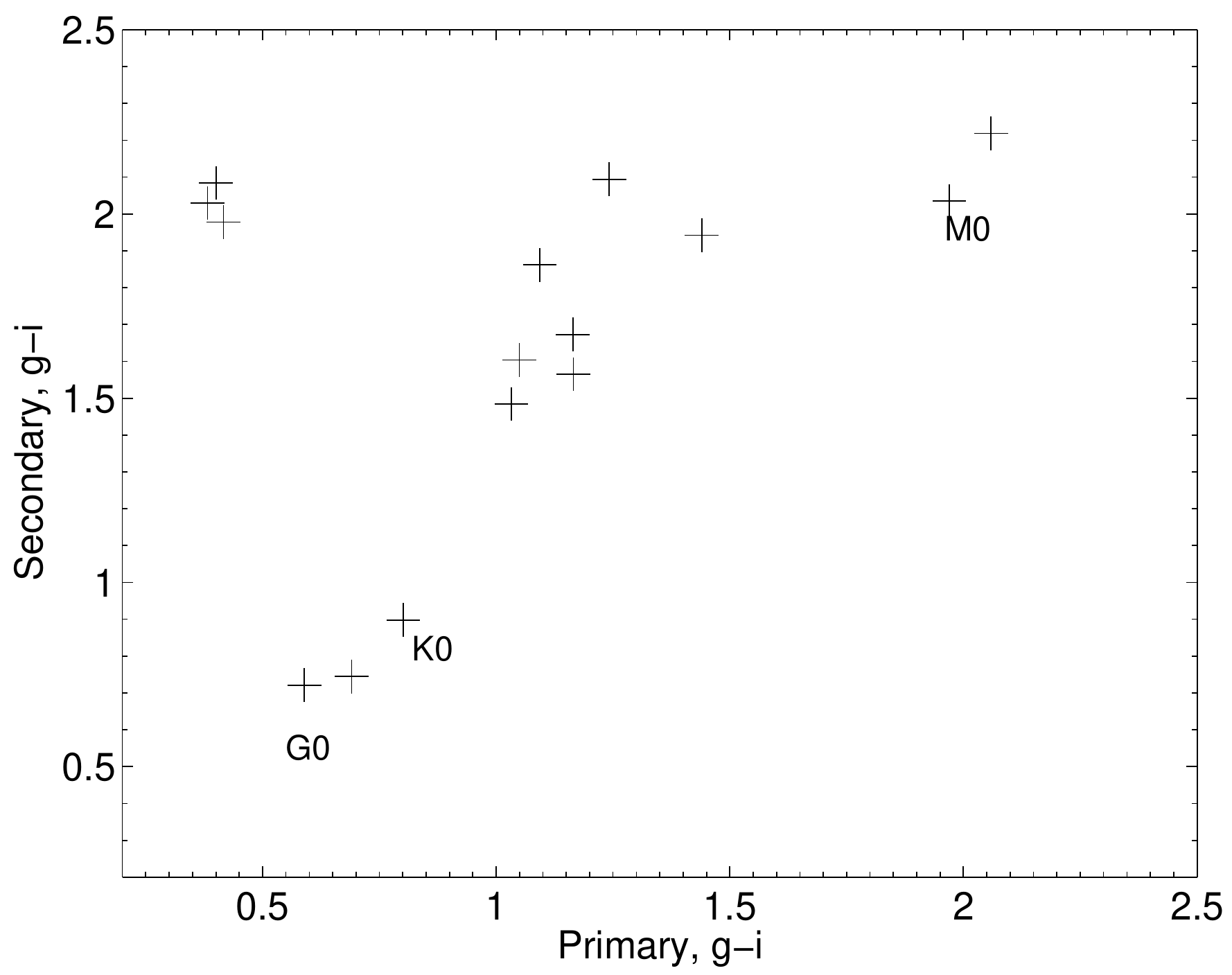}}}
\caption[]{\label{col}A colour-colour plot of the primary and secondary 
for each binary is shown. Spectral types from \cite{Covey} are indicated.}
\end{figure}
\section{Discussion and Conclusions}
\label{sec:end}
In this paper we have described a search for wide halo binaries in 
SDSS Stripe 82. Our main result is Table \ref{tab_cand}, which  provides the first set of high 
probability wide halo binary candidates observed 
in SDSS colours. As the projected separations lie between roughly 
0.01 and 0.25 pc, the new sample serves as a useful addition to the CG04 
sample of wide binaries. Over the interval $3\arcsec$ and $40\arcsec$ 
the sample obtained with 
the proper motion and photometric constraints is clean and 
likely free from selection effects as a 
function of $\Delta \theta$, so we can use it to 
find the power law that best characterizes the observed 
angular separation function. Our best fit leads to a power law index 
of -1.25$\pm0.4$. This is not as steep as the power law found in CG04 
but they are consistent within the uncertainty of the value of the fitted 
index. The deficit of pairs just outside $40\arcsec$ would appear to 
be inconsistent with a flatter power law unless it steepens, for example 
due to the action of compact dark matter bodies. However, for the moment 
this must remain a speculation, as the small size of our 
new sample relative to CG04 precludes any improvements on current 
dark matter constraints from wide halo binaries. 

Future surveys though, such as Pan-STARRS \citep{pan_stars} Gaia or the Large Synoptic Survey Telescope \citep[][]{LSST} will enable vaster samples of wide binaries to be selected with the aid of more stringent proper motion and trigonometric or photo-parallax constraints. For example, 
the 3.5 year Pan-STARRS PS1 3$\pi$ survey, which is due to  
begin this year, will observe 3/4 of the sky in the 5 stellar filters ($grizy$) 
with a predicted proper motion accuracy of 1.2 mas/yr \citep{pan_star2}. If we simply scale up the numbers  
found in Stripe 82 by this increased sky coverage, then the PS1 survey  
will yield at least 2000 wide halo binaries over an order of magnitude 
greater than than CG04 sample size. Such a sample would be  
hugely influential for advancing our understanding of these 
objects and placing robust constraints on the nature of dark matter. 

In the meantime spectroscopic observations of our new binary candidates 
for radial velocities and metalicites could be scientifically fruitful. 
As Figure~\ref{col} shows, many of the binaries 
span a wide range in spectral type and are therefore of 
particular importance in testing and constraining 
SDSS subdwarf photo-parallax relations. Indeed we pointed out 
that the scatter in the $\delta$ values in our sample is larger than expected 
for the I08 photo-parallax relation for subdwarfs 
stars. If this is due to unresolved triplets, spectroscopic observations 
could uncover evidence for this. In any case metalicities for the new sample 
would allow more complicated 
colour-magnitude relations, that change systematically with 
metalicity, to be tested.

With radial velocities from spectra, 
it will be possible to analyse the Galactic orbit of the objects, 
allowing one to address questions such 
as whether there is a trend in binary separation and the Galactic 
orbit averaged dark matter density. 
Another application that could be carried 
out, following spectroscopic observations of the sample, is to search for associations of wide binaries in integral 
of motion and metalicity parameter space, testing the 
idea put forward by \cite{Allen2}
that wide halo binaries may have survived disruption in the Galaxy on  
account of being recently accreted from disrupted satellites. 

\begin{figure}
\rotatebox{0}{\scalebox{.35}[.35]{\includegraphics{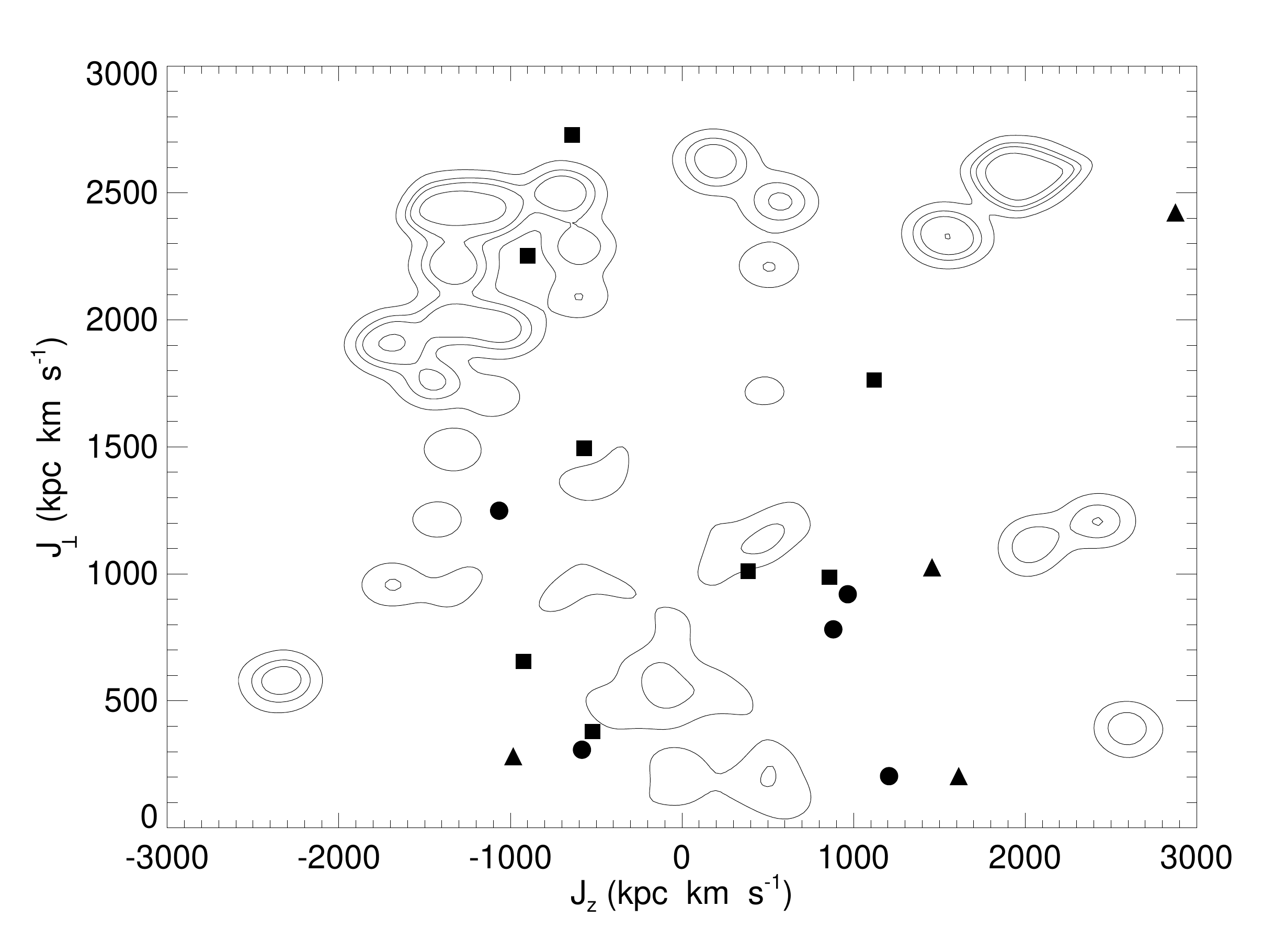}}}
\caption[]{\label{integrals}
The components of angular momentum perpendicular (J$_{\perp}$) and  
parallel (J$_{\textrm{z}}$) to the symmetry axis of the Galactic plane for 
the new  binaries with spectra from our sample (circles), along with the CG04  
wide halo binaries (projected separation$>0.01$ pc) with radial velocities measurements of  
both components (triangles) and those with radial  
velocities for just one component (squares). Also included in this figure are the contours  
from \cite{ms_hkin} showing the 1, 3, 5, 10 $\sigma$ over-densities in this space for halo  
subdwarfs within 2.5 kpc. Typical errors in (J$_{\textrm{z}}$, J$_{\perp}$) are around  
200-250 kpc.km.$s^{-1}$. Radial velocities for the CG04 wide binaries 
were obtained from the compilation in \cite{Quinn} and by searching 
in SIMBAD.} 
\end{figure}

We give a foretaste of such a test using the five new 
binaries with spectra from our sample, 
along with the CG04 wide halo binaries with projected 
separations larger than 0.01 pc and which radial velocities 
have been measured for one or both pairs.  
This is given in Figure \ref{integrals}, which shows the 
components of the angular momentum perpendicular and parallel 
to the symmetry axis of the Galactic plane, J$_{\perp}$ and 
J$_{\textrm{z}}$ in units of kpc.km.s$^{-1}$ respectively. 
The use of these two components 
is a common approach since they are adiabatic invariants 
in a spherical potential. Also included in this Figure are 
the contours from \citet[][ see figure 6]{ms_hkin} showing 
over-densities in this space for halo subdwarfs within 2.5 kpc. The  
over-density located around (J$_{\textrm{z}} \approx -1500$, J$_{\perp} \approx 2250$)  
is the potential accretion remnant identified by \cite{Helmi99}
and we can see that a couple of the wide binaries are possibly  
associated to this feature. 
None of the other binaries appear to be associated to any over-densities of more than 3 sigma, although it is worth noting the  
possible clumping amongst the binaries themselves. 

As more radial velocities  
become available it will be interesting to see how the distribution of 
wide binaries take shapes in this plot. Pockets of tight associations 
would be evidence in favour of any scheme in which the formation of 
wide binaries is associated with the disruption of a large stellar system.
Although some care may need to be taken in interpreting possible clumps on account of the complicated selection effects that have gone into defining the sample, 
we do not expect this application or indeed the other ones we have proposed for our new wide binary sample, to be seriously effected by these. 
\begin{table*}
\begin{small}\begin{tabular}{c|c|c|c|c|c|c|c|c|c|c|c|}
\hline
SDSS Obj$_{1}$&SDSS Obj$_{2}$&$\mu_{\alpha}cos(\delta)_{1}$&$\mu_{\alpha}cos(\delta)_{2}$&
$\mu_{\delta}$ $_{1}$&$\mu_{\delta}$ $_{2}$&$r_{1}$&$r_{2}$&$g-i$ $_{1}$&$g-i$ $_{2}$&$\Delta \theta$&Proj Sep\\ 
& & [mas/yr]&[mas/yr]&[mas/yr]&[mas/yr]& & & & & [arcsec] &[pc]\\
\hline
J002255.45+005427.6&J002255.81+005423.7&69.21&70.98&-17.01&-19.55&14.20&19.27&0.42&1.98&6.69&0.030\\      
J010740.28-004732.4&J010740.33-004725.2&-37.80&-37.86&-64.05&-64.35&15.65&15.30&0.72&0.59&7.20&0.029\\    
J011447.79-002339.0&J011448.38-002331.2&40.73&39.25&-14.88&-14.56&18.47&19.00&1.05&1.60&11.73&0.098\\     
J013335.18-002552.3&J013335.26-002549.8&35.15&36.06&-31.71&-27.79&20.64&20.87&1.97&2.04&2.83&0.029\\      
J014026.17-003229.8&J014026.33-003231.4&74.84&73.76&-3.47&-1.41&19.44&17.50&1.86&1.09&2.95&0.018\\        
J030126.09-004049.2&J030126.10-004053.3&105.83&107.99&5.35&0.86&16.59&19.43&1.24&2.09&4.08&0.016\\        
J212546.23-002939.1&J212546.46-002939.0&8.85&6.71&-51.29&-48.40&18.93&17.89&1.67&1.16&3.24&0.021\\        
J213457.20+002505.6&J213457.58+002504.9&-0.05&2.33&-51.61&-50.29&14.90&19.62&0.40&2.08&5.44&0.029\\       
J220549.2+002738.2&J220551.43+002731.3&-14.60&-17.76&-38.62&-41.49&15.32&19.93&0.38&2.03&34.18&0.230\\   
J221549.22+005737.0&J221549.63+005732.9&-5.45&-3.01&-142.00&-142.22&18.16&17.49&2.22&2.06&7.38&0.015\\    
J222937.88+000503.3&J222938.02+000503.7&-34.57&-36.73&-88.64&-86.43&17.16&16.44&1.57&1.17&2.19&0.007\\    
J223659.49+010744.4&J223700.42+010725.0&55.70&55.77&-54.73&-54.78&15.95&16.28&0.69&0.74&23.91&0.113\\     
J224154.05-003736.1&J224154.99-003734.3&4.31&4.43&-40.55&-42.39&18.73&20.21&1.44&1.94&14.16&0.112\\       
J224515.39-005706.5&J224517.43-005709.2&25.32&25.59&-38.00&-38.54&17.18&17.41&0.90&0.80&30.71&0.210\\     
J235153.62+000240.8&J235153.91+000237.5&47.43&45.65&-94.63&-93.58&17.34&16.71&1.48&1.03&5.73&0.023\\      
 & & & & & & & & & & &\\
J024131.67-010959.3& J024131.70-010954.7&2.16& -0.13 &-54.46 &-51.83&21.18&17.27&2.12&0.97&4.54&0.035\\
\hline
\end{tabular}
\end{small}
\caption[]{\label{tab_cand} Properties of the Stripe 82 candidate 
wide halo binaries. The last column lists 
the projected separation of the binary computed using the mean of the 
photo-parallax distances to each component of the binary, taking the metalicity 
to be -1.5 dex. The last entry 
failed to meet the photo-parallax constraint but is likely to be a binary (See 
discussion at end of Section~\ref{sec:search}).}
\end{table*}

\begin{table}
\begin{small}\begin{tabular}{c|c|c|}
\hline
SDSS Obj&Radial Velocity&[Fe/H]\\
&km/s&dex\\\hline\\
J010740.28-004732.4&9.5$\pm$2.0&-1.81$\pm$0.07\\
J030126.09-004049.2&142$\pm$2& -1.35$\pm$0.22\\
J221549.63+005732.9&-108$\pm$2&NA\\
J224515.39-005706.5&-244$\pm$3&-1.48$\pm$0.08\\
J235153.62+000240.8&-75$\pm$4&-2.17$\pm$0.17\\
\hline
\end{tabular}
\end{small}
\caption[]{\label{tab_2} 
Members of the Stripe 82 candidate 
wide halo binaries for which there are SDSS 
spectra. Heliocentric radial velocity and 
if available, metalicity are listed. The measurement errors 
are those given in the SDSS database.}
\end{table}

\section*{Acknowledgments}
We thank the referee for a quick 
and thoughtful response.
We are grateful to V. Belokurov and M. Wilkinson for suggesting 
this project. We would like to thank V. Belokurov for
many helpful discussions. DPQ acknowleges 
support from a Isbel Fletcher Studentship, a National University of Ireland 
Traveling Studentship, an Isaac Newton Studentship and Trinity College. 
MCS acknowledges support from the STFC-funded
``Galaxy Formation and Evolution'' program at the Institute of Astronomy.

\end{document}